\documentclass[final,5p,times,twocolumn]{elsarticle}
\makeatletter
\def\ps@pprintTitle{%
  \let\@oddhead\@empty
  \let\@evenhead\@empty
  \let\@oddfoot\@empty
  \let\@evenfoot\@empty}
\makeatother

\usepackage{amssymb}
\usepackage{amsmath}
\usepackage{siunitx}
\usepackage{hyperref}
\usepackage{orcidlink}
\usepackage{placeins}

\usepackage{comment}

\usepackage{lineno}
\journal{Nucl. Instr. Meth. A}

\begin{document}

\begin{frontmatter}

\title{Picosecond laser test unit for photosensor characterization\\ at ambient and low temperatures\tnoteref{t1}}
\tnotetext[t1]{This work was supported by the Deutsche Forschungsgemeinschaft~(DFG, German Research Foundation) through the Collaborative Research Center~SFB1258 ``Neutrinos and Dark Matter in Astro- and Particle Physics“, and the Excellence Clusters ORIGINS~(EXC 2094) and PRISMA$^+$~(EXC~2118).}

\author[a]{Matthias Raphael Stock\corref{cor1} \orcidlink{0000-0002-5963-7431}}%\fnref{fn1}
\ead{raphael.stock@tum.de}
\cortext[cor1]{Corresponding author}
\author[a,b,c]{Hans Th. J. Steiger \orcidlink{0000-0003-3323-1377}}
\author[a]{Ulrike Fahrendholz \orcidlink{0009-0002-2128-2214}}
\author[a]{Luca Schweizer \orcidlink{0009-0001-2334-9526}} 
\author[a]{Lothar Oberauer \orcidlink{0009-0008-6238-208X}}

%% Author affiliation
\affiliation[a]{organization={Technical University of Munich, TUM School of Natural Sciences, Department of Physics},%Department and Organization
            addressline={James-Franck-Str. 1}, 
            %postcode={}, 
            city={85748 Garching},%state={},
            country={Germany}}

\affiliation[b]{organization={Johannes Gutenberg University Mainz, Cluster of Excellence PRISMA$^+$},%Department and Organization
            addressline={Staudingerweg 9}, 
            %postcode={55128}, 
            city={55128 Mainz}, %state={},
            country={Germany}}
\affiliation[c]{organization={Johannes Gutenberg University Mainz, Institute of Physics}, %Department and Organization
            addressline={Staudingerweg 7}, 
            %postcode={55128}, 
            city={55128 Mainz},%state={},
            country={Germany}}

\begin{abstract}
Accurate single photoelectron~(SPE) characterization of photosensors is essential for controlling systematic uncertainties in low-light neutrino and dark matter detectors.
We present a compact laboratory setup for the characterization of photosensors under controlled, low-light conditions.
Specifically, we demonstrate its use with photomultiplier tubes~(PMTs) operated at the~SPE-level, using picosecond laser pulses and waveform digitization to determine key~PMT properties.
Measurements as a function of supply voltage and temperature~($-50^\circ$C to~$+20^\circ$C) are performed on ET~Enterprises~9821(Q)B tubes and a Hamamatsu~R9980 assembly, which show exponential gain-voltage behavior and device-to-device variation.
Cooling increases the gain by~$\sim 0.1\,\%/^\circ$C, while the transit time spread~(TTS) and peak-to-valley ratio~(P/V) exhibit no clear temperature dependence.~TTS decreases with voltage.
Late pulses remain at the percent level and prepulses at the sub-percent level.
Cable length affects both apparent gain and~TTS.
A model-independent, data-driven self-convolution method is introduced to quantify double photoelectron contributions from pulse charge spectra.
The procedures provide a reproducible, practice-oriented reference for SPE-level PMT characterization and can be extended to other photosensor types.
\end{abstract}

\begin{comment}
%%Graphical abstract
%\begin{graphicalabstract}
%\includegraphics{grabs}
%\end{graphicalabstract}

%%Research highlights (3 to 5)
\begin{highlights}
\item Compact picosecond laser test bench enabling reproducible~SPE-level charge and timing characterization of~PMTs.

\item Unified waveform-level analysis of gain,~P/V,~TTS, and prepulse/late pulse populations across multiple~PMT types.

\item Model-independent determination of double photoelectron contributions from pulse charge spectra via a data-driven self-convolution method.

\item Temperature-dependent characterization from~$-50^\circ$C to~$+20^\circ$C with quantified effects on gain and negligible impact on~TTS and~P/V.

\item Demonstrated influence of cable length on apparent gain and timing, emphasizing the importance of consistent electrical routing.
\end{highlights}
\end{comment}

%% Keywords

\end{frontmatter}

%% ---------------- Main text ----------------

\section{Introduction}\label{sec:intro}
Modern low-background neutrino and dark matter experiments rely on large arrays of photosensors to detect faint scintillation and Cherenkov light. 
Accurate knowledge of the single photoelectron~(SPE) response, gain, and timing is required to control systematic uncertainties in reconstructed energy, position, and event topology.
Photomultiplier tubes~(PMTs) are widely used in liquid scintillator and Cherenkov detectors because they combine~SPE sensitivity with nanosecond-scale timing. 
Dedicated test stands have supported~PMT qualification and calibration for large experiments such as for Borexino~\cite{BRIGATTI2005521,PMT_SmiLomRan2004}, Double Chooz~\cite{QualiPMTs_DC_2011,TTandCh_SPE_R7081_PMTs_2012}, XENON1T~\cite{QualiPMTs_XENON_2017} and JUNO~\cite{JUNO}, typically emphasizing high-throughput testing, robust calibration procedures, and long-term stability.

In parallel, more compact and flexible laboratory setups are needed for detailed~PMT and photosensor studies in R\&D environments. 
Comprehensive photosensor studies for large liquid scintillator detectors, for example in the context of the proposed~LENA observatory, have highlighted this need~\cite{PhDTippmann}. 
Building on this experience, we employ a compact, table-top bench with picosecond laser excitation, controlled illumination, and synchronized data acquisition~(DAQ) for photosensor characterization, and demonstrate its performance with~PMTs.

From the recorded waveforms, gain and peak-to-valley ratio~(P/V) are obtained from charge spectra, while transit time spread~(TTS) and prepulse/late pulse populations are extracted from transit time distributions. 
Optional temperature scans in a climate chamber extend the characterization to controlled low and ambient temperatures. 
Parameters such as absolute transit time, afterpulse probability, high-light linearity, and photon detection efficiency are outside the present scope, but the methods can be extended to include them. 
For example, operating an additional reference photosensor, such as a silicon photomultiplier~(SiPM) viewing the same laser pulses, would in principle allow one to determine the~PMT transit time from relative timing and to cross-check high-intensity calibration and linearity.

While many aspects of~PMT characterization are well established, this work provides a concise, practice-oriented reference that combines a compact setup with waveform-based analysis procedures applicable to a wide range of~PMT types. 
In addition to the standard charge and timing observables, a model-independent method is introduced to quantify the double photoelectron~(DPE) contribution directly from the pulse charge spectrum.
The characterization procedures developed here provide the basis for describing the~PMT responses in our liquid scintillator~(LS)~R\&D program and are used to verify and validate the performance of the~LS test setup that underlies the measurements reported in~\cite{WbLS, SlowLS, TAUP2023}.

\section{Experimental setup}\label{sec:PMTUnitExpSetup}
\begin{figure}[htb]
    \centering
    \includegraphics[width=\linewidth]{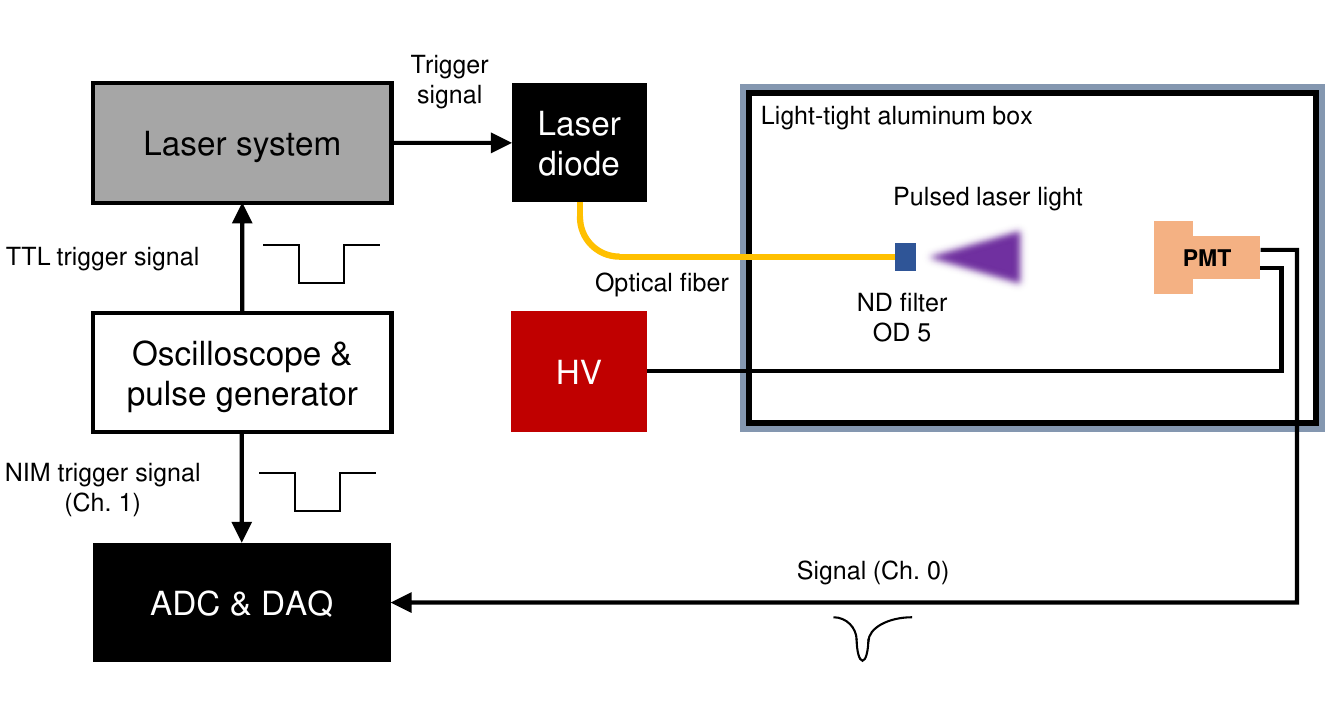}
    \caption{Schematic of the experimental setup for photosensor characterization. 
    A~TTL signal triggers the laser, while a synchronized~NIM pulse triggers the digitizer. 
    The attenuated laser beam illuminates the PMT inside a light-tight enclosure.}
    \label{fig:PMTtestUnit}
\end{figure}

The test and characterization unit enables reproducible measurements of photosensor responses at the~SPE-level under controlled laboratory conditions. 
Figure~\ref{fig:PMTtestUnit} outlines the system, which consists of a picosecond pulsed laser, a light-tight enclosure or climate chamber, and a synchronized acquisition chain. 
The main components are summarized below.
 
\paragraph{Laser system} 
Excitation light is provided by a PILAS~DX diode laser head with a PILAS~D2 driver~(NKT Photonics) at a wavelength of~407\,nm. 
Typical pulse widths are~21–34\,ps~FWHM with timing jitter below~3\,ps, and repetition rates up to~40\,MHz. 
The output is coupled to a single-mode fiber~(SM400, Thorlabs) and a neutral density filter of optical density~5 attenuates the beam by a factor of~$10^5$.
The laser intensity is further adjusted so that about~5–10\,\% of triggers yield detected~PMT responses.
This operating point corresponds to~SPE-level, low-light conditions while maintaining a favorable balance between real signals and dark noise.

The effective temporal width of the laser pulses was verified using a KETEK PM3325-WB~SiPM placed~3.3\,cm from the fiber output, operated at~29.7\,V. 
The time difference between~SiPM signals and the laser trigger follows a Gaussian distribution with FWHM~$=(12.4\pm0.1)$\,ps~\cite{PhD_MRStock2025}, well below the~PMT~TTS values studied here~($\gtrsim 180$\,ps), so no deconvolution of transit time spectra is required.

\begin{figure}[htb]
    \centering
    \includegraphics[width=\linewidth]{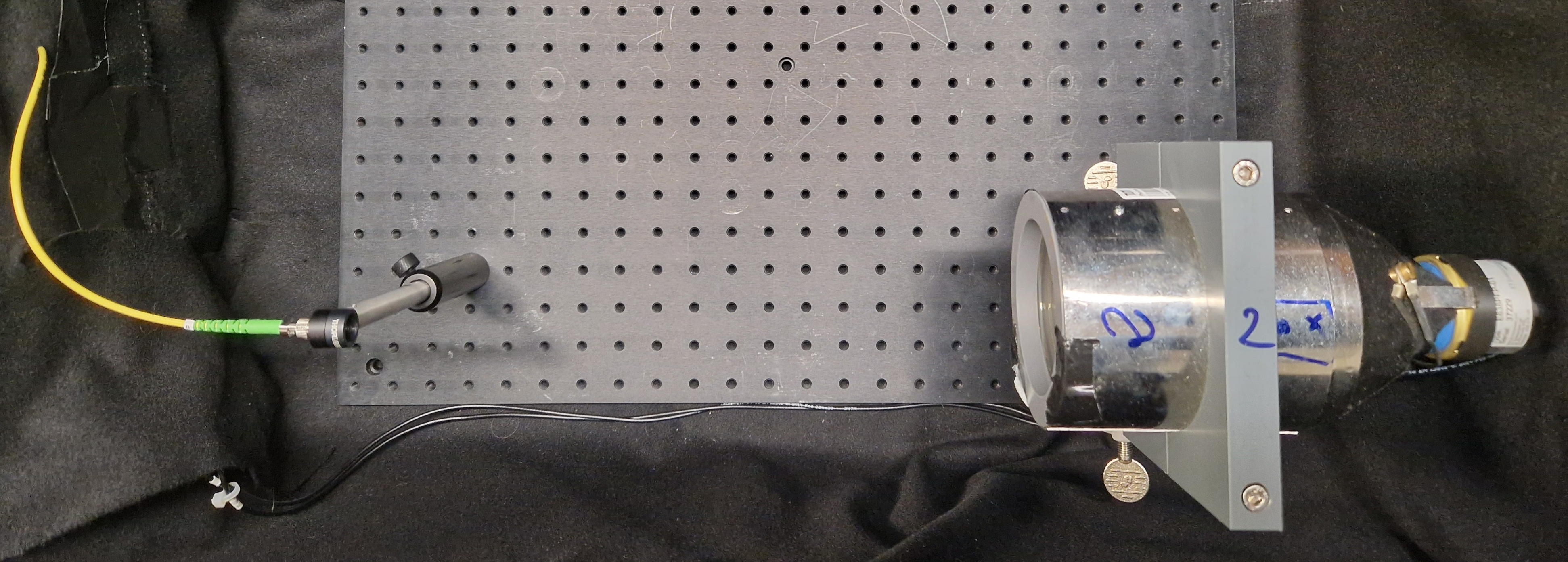}
    \caption{Photograph of the~PMT inside the light-tight enclosure. 
    The fiber output is positioned about~35\,cm from the photocathode, ensuring full illumination.}
    \label{fig:Innen}
\end{figure}

\paragraph{Light-tight enclosure and climate chamber}
The~PMT is mounted either in a light-tight, blackened aluminum box~(approximately~$70 \times 54 \times 40\,\mathrm{cm}^3$ in size) or in a Binder~MKT~115 climate chamber. 
Both provide optical isolation and act as Faraday cages.
The fiber and cables enter through light-tight feedthroughs, and the fiber is positioned about~35\,cm from the photocathode, as shown in figure~\ref{fig:Innen}, ensuring full coverage of the active area. 
Under far-field geometry, the laser spot size increases approximately linearly with the fiber-to-photocathode distance.
Although the irradiance is not completely uniform across the photocathode, this geometry is sufficient for~SPE-level characterization of~3-inch~PMTs. 
A detailed study of the illumination profile can be found in~\cite{PhD_MRStock2025}. 
For temperature studies, the~PMT, fiber and signal cables are placed inside the climate chamber, operated within~$-50^\circ$C to~$+20^\circ$C.
After each temperature change, data taking begins only after sufficient equilibration time of all components.

\paragraph{Electronics, trigger and synchronization} 
The~PMTs are powered by a CAEN N1470 high-voltage supply with~1\,V adjustment precision. 
A Rohde \& Schwarz RTO2004 oscilloscope generates two synchronized logic signals. 
A~TTL pulse triggers the laser, while a fast-negative~NIM pulse triggers data acquisition. 
The~NIM trigger can be phase-shifted so that both the~PMT pulse and trigger lie within the selected~DAQ time window.

The laser repetition rate is typically~20\,kHz or lower (e.g.,~10\,kHz), corresponding to pulse spacings of 50–100\,µs. 
The~DAQ time window is~100\,ns, which is more than five orders of magnitude shorter than the interval between consecutive laser pulses. 
Consequently, each recorded waveform contains the PMT response to at most one laser pulse, preventing overlap of independent events within a single acquisition time window.

The timing precision of the waveform generator was verified by generating synchronized ~TTL and~NIM logic signals from the~RTO2004 oscilloscope and recording the waveforms at~10\,kHz using the internal~10\,GS/s~ADC, with interpolation applied to improve resolution. 
The resulting time difference distribution exhibits a dominant Gaussian component with FWHM~$=(8.6 \pm 0.2)$\,ps and an overall width of~33\,ps~\cite{PhD_MRStock2025}, both negligible compared to typical PMT transit time spreads.

\paragraph{Data acquisition and processing}  
Waveforms are digitized without thresholding using an Acqiris~DC282 compactPCI digitizer~(U1065A) with up to~8\,GS/s sampling rate,~10-bit voltage resolution, and~2\,GHz analog bandwidth. 
For multi-channel operation, a rate of~2\,GS/s rate is used, corresponding to~$\Delta t = 0.5$\,ns/bin. 
Signals are back-terminated at~50\,$\Omega$, and selectable input ranges from~50\,mV to~5\,V accommodate different pulse amplitudes. 

Data are stored in binary format and converted into~ROOT~\cite{ROOT} files using a dedicated~C++ software framework originally developed for the positronium lifetime determination experiment~\cite{SCHWARZ201964}. 
The software reconstructs waveforms and extracts parameters such as baseline parameters, pulse height, integrated charge, and timing information. 
A graphical interface supports validation and inspection of individual events.

For diagnostics, the~RTO2004 oscilloscope operates in parallel with the digitizer. 
If necessary, it can also serve as the primary acquisition device with up to~10\,GS/s,~600\,MHz bandwidth and~8-16\,bit resolution. 
However, all quantitative results presented here are based on Acqiris data.

\section{Pulse characterization}\label{subsec:PulseCharacterization}
\begin{figure}[htb!]
    \centering
    \includegraphics[width=1\linewidth]{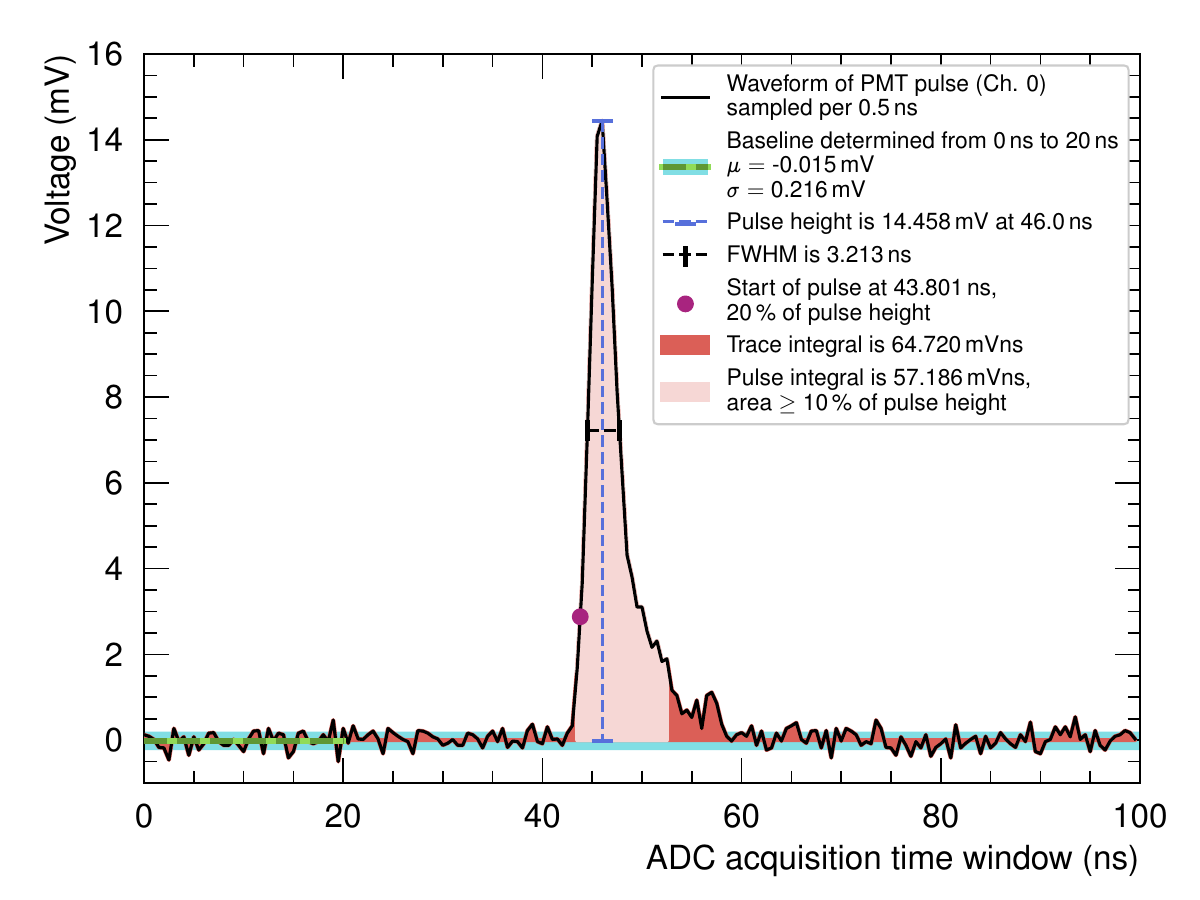}
    \caption{Example waveform from a~3-{inch}~9821B~PMT at the~SPE-level. 
    The baseline mean~(green) and its standard deviation~(light blue band) quantify the noise. 
    The dashed blue line marks the pulse height, the dashed black line the~FWHM, the purple dot the start time from constant fraction timing~(20\,\%), and the shaded red and rose areas the trace and pulse integrals, respectively.}
    \label{fig:PulseSketch}
\end{figure}

Several parameters are defined to describe~PMT signal waveforms. 
Since the negative~PMT signals are inverted, all pulses appear positive, as shown in figure~\ref{fig:PulseSketch}.
The main pulse descriptors are:
\begin{itemize}
    \item \textbf{Baseline:} estimated from an initial portion of the waveform, such as~20\,ns. 
    The mean~$\mu$ and standard deviation~$\sigma$ quantify the noise level.
    \item \textbf{Pulse height:} maximum amplitude relative to the baseline,~$U_\text{max} - \mu$.
    \item \textbf{Start time:} determined with a constant fraction algorithm, e.g., using~20\,\% of the pulse height, to reduce time walk.
    \item \textbf{Number of pulses:} identified using a threshold (typically~$2\,\sigma$ of the baseline) and minimum width~(5\,ns) to suppress spurious noise triggers.
    \item \textbf{Trace integral:} area under the entire baseline-corrected waveform, providing a measure of the total collected charge.  This quantity includes noise outside the pulse region and thus depends on the~DAQ time window.
    
    \item \textbf{Pulse integral:} area under the pulse above~10\,\% of its pulse height relative to the baseline. 
    This definition excludes baseline contributions outside the pulse and is therefore less sensitive to the chosen~DAQ time window.
\end{itemize}
These definitions provide a consistent basis for analyzing PMT gain, timing properties, and noise contributions.

%--------------------------------------------------
\section{Data analysis and methods}\label{sec:analysis}
The following analysis procedures are illustrated using example data from a~3-inch 9821B~PMT operated at~1800\,V supply voltage.
All results shown here are representative of the general methodology applied to all tubes.
Each measurement typically contains more than one million events, allowing precise determination of the distributions of relevant pulse descriptors. 
Histogram bin widths are chosen as small as permitted by the~ADC sampling granularity, maximizing resolution while avoiding digitization artifacts.
All analysis is performed in \texttt{Python}~3.8~\cite{python} using \texttt{numpy}~1.24.3~\cite{numpy},~\texttt{scipy}~1.10.1~\cite{scipy}, \texttt{matplotlib}~3.72~\cite{matplotlib}, \texttt{uproot}~5.3.13~\cite{uproot}, \texttt{iminuit}~2.29.1~\cite{iminuit}, \texttt{SymPy}~1.13.3~\cite{Sympy} and \texttt{pandas}~2.0.3~\cite{pandas}.
The basic event selection and preprocessing steps are as follows:
\begin{enumerate}
    \item The baseline mean and baseline standard deviation spectra are modeled with one to three Gaussian components using an extended binned maximum likelihood fit (via \texttt{iminuit}~\cite{iminuit}). 
    The preferred model is chosen via the reduced~$\chi^2$, and~$'3\,\sigma'$ selection cuts are typically applied to suppress outliers. 
    
    \item For noise studies of the trace charge spectrum, waveforms with zero or one identified pulse are retained. 
    In all other cases, such as transit time spectrum studies, exactly one pulse per waveform is retained.
    
    \item The pulse height distribution is approximated using a Gaussian kernel density estimation~(KDE) implemented in \texttt{scipy.stats.gaussian\_kde}~\cite{scipy}.
    This~KDE method provides a smooth, binning-independent representation of the distribution.
    Its bandwidth is tuned to avoid artificial structure while retaining physical features.
    Depending on the pulse identifier requirement, either a distinct valley above the noise peak is visible or the noise population is fully suppressed.
    If the valley is visible, its position is taken from the~KDE and used as the lower pulse height threshold to reject noise events.
    If the valley is absent, a fixed lower threshold is used,~e.g., the voltage at which the~KDE reaches~50\,\% of its maximum.
    Saturated events close to the~ADC voltage limit are identified and removed.
    Except for dedicated studies of the trace charge spectrum, both the pulse height noise and saturated components are excluded from the analysis.
\end{enumerate}
After these basic data quality selections, the data are ready for extraction of trace charge spectra, pulse charge spectra, and transit time spectra~\cite{PhD_MRStock2025}.
The following subsections describe these procedures and corresponding statistical models.

\subsection{Trace charge spectrum}
Trace integrals are used to select~SPE events, extract the~SPE mean charge~$\mu_{\text{SPE}}$, and determine the~PMT gain~$G$. 
The integrals are recorded in~mVns and converted to charge~(pC) using~$q = {U \times \Delta t}/{R}$ with the back-termination resistance~$R=50\,\Omega$.
Bins containing saturated events are excluded from the fit by scanning downward from the maximum charge bin. 
To describe the trace charge distribution, we adopt and extend the model of~\cite{Diwan2020}.
At low-light levels, the number of photoelectrons~(PEs) per trigger~$k$ follows a Poisson distribution with mean~$\lambda$. 
The probability of observing~$k$\,PEs is then,~$P(k \mid \lambda) = {\lambda^k \, \mathrm{e}^{-\lambda}} / {k!}$, which serves as the statistical weighting factor for each~$k$\,PE component. 

In the zero~PE case~($k=0$), the background charge distribution is the convolution of two components. 
The first arises from baseline current fluctuations (no photoelectron emission) and defines the so-called pedestal modeled as a Gaussian distribution. 
The corresponding probability density function~(PDF) is
\begin{equation}
G_{\text{PDF}}(q\,|\,q_0,\sigma_0) = \frac{1}{\sqrt{2\pi}\sigma_0}\exp\left[-\frac{(q - q_0)^2}{2\sigma_0^2}\right],
\end{equation}
where~$q$ is the charge variable,~$q_0$ is the mean pedestal charge, and~\mbox{$\sigma_0 > 0$} is the noise standard deviation~\cite{Diwan2020}. 
The second background component comes from discrete processes (thermionic dark pulses or under-amplified signals~\cite{QualiPMTs_DC_2011, QualiPMTs_XENON_2017, Diwan2020}).
These processes include~PEs missing the first dynode and~PE emission from focusing electrodes or dynodes, empirically described by
\begin{equation}
D_{\text{PDF}}(q \, | \, w_0, \tau_0) = (1 - w_0) \, \delta(q) + w_0 \, \frac{\Theta(q)}{\tau_0}\exp \left(-\frac{q}{\tau_0}\right),
\end{equation}
where~$w_0 \in (0, 1)$ is the dark pulse probability,~$\delta(q)$ is the Dirac delta function,~$\Theta(q)$ is the Heaviside step function and~$\tau_0 > 0$ is the exponential scale parameter. 

The convolution of a Gaussian distribution and an exponential distribution results in the so-called 
exponentially modified Gaussian distribution~(EMG),
\begin{align}
EM&G_{\text{PDF}}(q \, | \, \mu_0, \sigma_0, \tau_0) \notag\\
= & \frac{1}{2 \tau_0} \exp\left(\frac{\sigma_0^2 + 2 \tau_0 \mu_0 - 2 \tau_0 q}{2 \tau_0^2}\right) \,\mathrm{erfc}\left(\frac{\sigma_0^2 + \tau_0 \mu_0 - \tau_0 q }{\sqrt{2}\,\sigma_0 \tau_0} \right),
\end{align}
where~$\mathrm{erfc}(x)$ is the complementary error function. 

In the~SPE case~($k = 1$),~PEs collected at the first dynode induce the emission of secondary electrons, which multiply further through the dynode structure. 
The amplification factor of this process is known as gain~$G$, whose fluctuation is mainly dominated by the first dynode.
Under the assumption of a sufficiently large first-dynode secondary-emission coefficient,~e.g.,~$> 4$, and a collection coefficient near unity, the~SPE charge distribution can be modeled as a Gaussian distribution~$G_{\text{PDF}}(q\, |\, q_1, \sigma_1)$ with mean charge~$q_1$ and standard deviation~$\sigma_1$~\cite{Diwan2020}.

For multiple~PE~($k \geq 2$), each~PE is assumed to undergo independent multiplication. 
Thus, the resulting charge distribution for~$k$\,PEs corresponds to the sum of the distributions for double~PE~(DPE), triple~PE~(TPE), and so on. 
Since the convolution of~$k$ independent Gaussians is Gaussian with mean and variance scaled accordingly, the~$k$\,PE distribution is given by~$G_{\text{PDF}}(q\, |\, kq_1, \sqrt{k}\sigma_1)$.
This accounts for the expected charge increase with~$k$ while incorporating statistical gain fluctuations.

Additionally, the signal charge components are influenced by noise as well and therefore, described by the convolution of the background charge component~$D_{\text{PDF}}$ and the~$k$\,PE distributions.

However, following~\cite{RDossi2000}, the signal components are here truncated below the pedestal mean~$q_0$ using the Heaviside step function~$\Theta(q - q_0)$ to better describe the data.
Therefore, a Heaviside modified Gaussian distribution is defined as
\begin{equation}
GH_{\text{PDF}}(q\, |\, q_0, \mu, \sigma) = \sqrt{\frac{2}{\pi \sigma^2}} \frac{\Theta(q - q_0)}{1 - \mathrm{erf}\left( \frac{q_0 - \mu}{\sqrt{2} \sigma}\right) } \exp{\left(-\frac{\left(q - \mu \right)^2}{2\sigma^2} \right)},
\end{equation}
where~$\mathrm{erf}(x)$ is the error function. 
Similarly, the Heaviside modified exponentially modified Gaussian distribution is given by: \begin{align}  EM&GH_{\text{PDF}}(q \, | \, q_0, \mu, \sigma, \tau_0) = \Theta(q - q_0) \, \times \notag \\ & \frac{\exp{\left(\frac{\sigma^2 + 2 \tau_0 \mu - 2 \tau_0 q}{2 \tau_0^2}\right)} \ \mathrm{erfc}{\left(\frac{\sigma^2 + \tau_0 \mu - \tau_0 q}{\sqrt{2} \sigma \tau_0}\right)}}{\tau_0 \left[1 + \mathrm{erf}{\left(\frac{\mu - q_0}{\sqrt{2} \sigma}\right)} + \exp{\left(\frac{\sigma^2 + 2 \tau_0 \mu - 2 \tau_0 q_0}{2 \tau_0^2}\right)} \ \mathrm{erfc}{\left(\frac{\sigma^2 + \tau_0 \mu - \tau_0 q_0}{\sqrt{2} \sigma \tau_0}\right)} \right]}. \end{align}

Finally, the full model for the trace charge spectrum is
\begin{align}\label{eq:TraceChargeModel}
	\text{Model}_{\text{Trace charge}}&\left( q\, |\, N_{\text{total}}, \lambda, w_0, q_0, \sigma_0, \tau_0, \mu_1, \sigma_1 \right) = \nonumber \\  
	N_{\text{Events}} \times \Biggl( \mathrm{e}^{-\lambda} \times & \, \left[ \left( 1 - w_0 \right) \times G_{\text{PDF}}\left( q\, |\, q_0, \sigma_0 \right) \right. \notag \\
    & \left. + \, w_0 \times EMG_{\text{PDF}}\left( q \, | \, q_0, \sigma_0, \tau_0 \right) \right] \nonumber \\
	+ \sum^{n}_{k = 1} \frac{\lambda^k \mathrm{e}^{-\lambda}}{k!} \times & \nonumber \\
	\Biggl[ \left( 1 - w_0 \right) \times & \, GH_{\text{PDF}}\left( q\, | \, q_0, k \mu_1 + q_0, \sqrt{k \sigma_1^2 + \sigma_0^2} \right) \nonumber \\
    + \, w_0 \, \times & \, EMGH_{\text{PDF}} \left( q \, | \, q_0, k\mu_1 + q_0, \sqrt{k \sigma_1^2 + \sigma_0^2}, \tau_0 \right) \Biggr] \Biggr),
\end{align}
where~$N_{\text{total}}$ is the total number of events in the charge spectrum, and~$n$ is the number of~PE components, typically sufficiently described by~2 to~3.
Table~\ref{tab:TraceChargeModel_Parameters} lists the model parameters including a short description.

\begin{figure}[htb!]
    \centering
    \includegraphics[width=\linewidth]{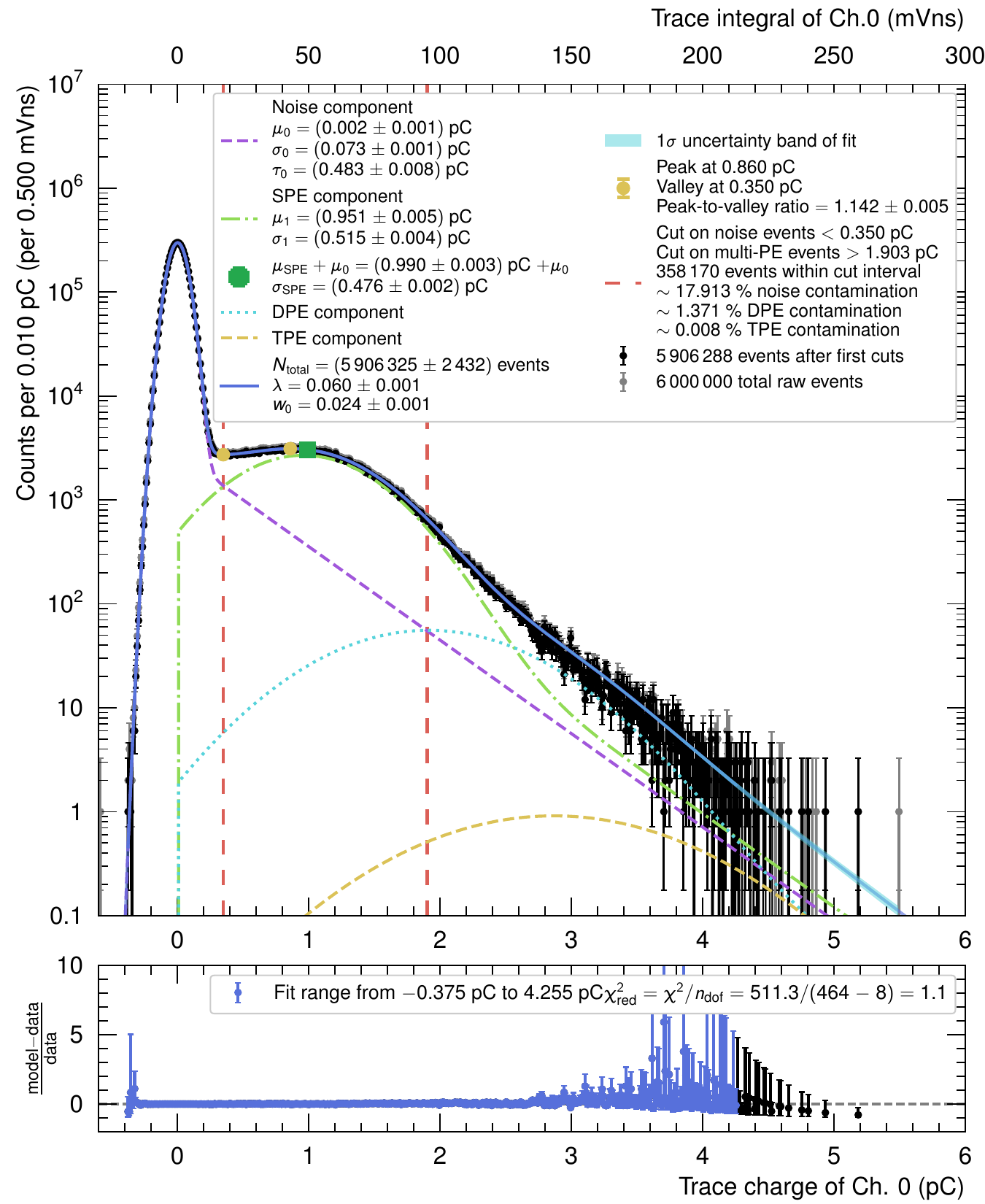}
    \caption{Example trace charge distribution at~SPE intensity from a~9821B~PMT, operated at~1800\,V supply voltage. The blue curve is the best-fit model. 
    Colored components indicate noise (purple dashed),~SPE (green dash-dotted),~DPE (cyan dotted), and~TPE (gold dashed) contributions. Vertical red dashed lines mark the~SPE selection window.
    The green marker denotes the~SPE mean~$\mu_{\text{SPE}}$.}
    \label{fig:TraceCharge}
\end{figure}

\begin{table}[htb!]
    \centering
        \begin{tabular}{lll}
        \hline%\hline
        Component & Parameter & Description \\
        \hline%\hline
        Fluctuating & $q_0$ & Mean of pedestal noise \\
        baseline current & $\sigma_0$ & Standard deviation of \\
         & & pedestal noise \vspace{0.5em} \\
        %\hline
        Dark pulses / & $\tau_0$ & Exponential scale parameter \\
        under-amplified & & of dark pulses \\
         & $w_0$ & Probability of dark pulses        \vspace{0.5em} \\
        %\hline
        SPE response & $\mu_1$ & pre-truncated SPE mean \\
        & $\sigma_1$ & pre-truncated SPE \\
        & & standard deviation         \vspace{0.5em}  \\
        %\hline
        Illumination & $\lambda$ & Mean number of \\
        & &  photoelectrons    \vspace{0.5em} \\
        %\hline
        Normalization & $N_{\text{total}}$ & Number of events \\
        \hline%\hline
        \end{tabular}
    \caption{Parameters of the trace charge spectrum model (see equation~\ref{eq:TraceChargeModel}).}\label{tab:TraceChargeModel_Parameters}
\end{table}
Fits are performed with an extended binned maximum likelihood method using the~\texttt{iminuit} library~\cite{iminuit}. 
Figure~\ref{fig:TraceCharge} shows a fit of the model, lists individual components and best-fit parameters. 
The lower panel highlights the residuals and indicates the data range used for fitting~(colored in blue), excluding saturated events.
The probability of dark pulses,~$w_0 \approx 2\,\%$, is relatively small.
The mean number of photoelectrons, here~$\lambda = 0.06$, can be used to derive the occupancy~$O_{\text{PMT}} = 1 - \exp(-\lambda) = 6\,\%$, implying that out of every hundred laser pulses, only six are detected by the PMT and confirming the low-intensity light~\cite{PMT_DEAP3600_2019}.
The peak in the~SPE regime~$P$ is identified as the local maximum of the fitted charge spectrum above the pedestal. 
The valley~$V$ is defined as the local minimum between pedestal and~$P$. 
Uncertainties of both peak and valley are determined from the~$1\sigma$ uncertainty band of the fit, which was derived from the covariance matrix of the fitted parameters.
By this, the peak-to-valley ratio~(P/V) can be determined, quantifying the separation power between noise and signal~\cite{QualiPMTs_XENON_2017}. 

An interval primarily containing~SPE events can be selected based on the fit curve, as shown in figure~\ref{fig:TraceCharge}. 
The lower bound is~$V$ when present and otherwise, the breadth of the pedestal is evaluated using the condition~$\mu_0 + 3 \sigma_0 > \mu_1$.
If the pedestal is broad, the minimum is set to~$\mu_0 + \mu_1$, otherwise to~$\mu_0+\mu_1/2$. 
The upper bound is chosen as~$\mu_0 + 2 \mu_1$ to limit multi-PE contamination. 
Integrating the model components over this interval yields contamination fractions from noise,~DPE, and~TPE.
Here, the~DPE contamination of~$\sim 1.4\,\%$ is small and~TPEs are negligible, while noise can dominate the residual contamination.

Since the signal charge distributions are cut off at~$q_0$, the true mean value of the~SPE component is actually shifted to the right, and its real standard deviation also deviates from the initial parameter. 
Using the central moment definition, the true mean of the~SPE component~$\mu_{\text{SPE}}$ is calculated as
\begin{align} 
\mu_{\text{SPE}} & = \int^{\infty}_{-\infty} q' \times GH_{\text{PDF}}(q'\, |\, q_0, \mu_1, \sigma_1) \; \mathrm{d}{q'} \notag \\ & = \mu_1 + \sqrt{\frac{2 \sigma_1^2}{\pi}} \frac{\exp{\left(-\frac{\left(\mu_1 - q_0\right)^2}{2 \sigma_1^2}\right)}}{1 + \mathrm{erf}\left(\frac{\mu_1 - q_0}{\sqrt{2} \sigma_1}\right)}. 
\end{align} 

This corrected mean value is highlighted in green in figure~\ref{fig:TraceCharge}. 
Similarly, the true standard deviation~$\sigma_{\text{SPE}}$ is computed.
Uncertainties of both values are determined using uncertainty propagation, based on the best-fit parameters and their statistical uncertainties. 
The partial derivatives required for this propagation were numerically computed using the~\texttt{Sympy} library~\cite{Sympy}. 

The gain is calculated by~$G = \mu_{\text{SPE}}/e$~\cite{PMT_Bellamy_1994}, where~$e$ is the electric charge.
Here, the gain is~$G = 5 \times 10^6$ at~1800\,V, which is in agreement with the datasheet~\cite{9821B_datasheet}.

\subsection{Pulse charge spectrum}
\begin{figure}[htb]
    \centering
    \includegraphics[width=1\linewidth]{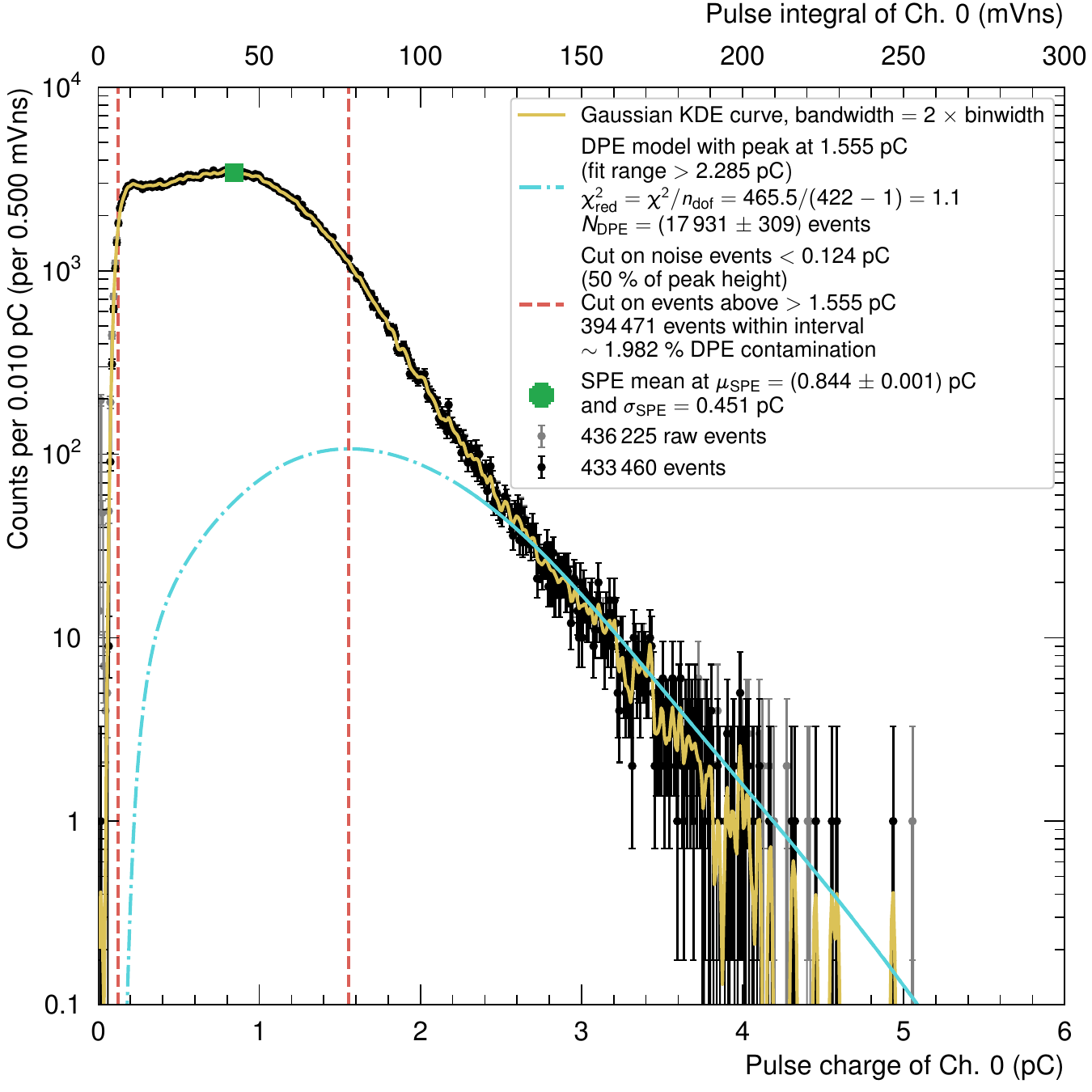}
    \caption{Example pulse charge spectrum at~SPE intensity for a~9821B~PMT operated at~1800\,V. 
    The gold curve is a Gaussian~KDE, and the green marker indicates the~SPE mean charge.
    Vertical red dashed lines denote the~SPE selection window.
    The cyan dash-dotted curve is a~DPE model fitted to the high-charge tail to estimate~DPE contamination within the~SPE window, while the solid cyan line indicates the data range used for the fit.}
    \label{fig:PulseCharge}
\end{figure}

The pulse integral data can also be used to define data quality cuts for selecting~SPE events, to extract the~SPE mean charge~$\mu_{\text{SPE}}$, and to compute the~PMT gain~$G$. 
Pulse integrals are converted to charge as above. 
Figure~\ref{fig:PulseCharge} shows an analyzed example spectrum, where the~DPE contamination in a proposed~SPE selection interval is indicated.
The method to estimate the~DPE distribution is explained in the following.

Assuming that most entries originate from~SPE pulses, the histogram is self-convolved to determine the position and shape of the~DPE distribution.
This self-convolution approach provides a model-independent estimate of the~DPE component, avoiding assumptions about the underlying~SPE shape and requiring only the measured pulse charge distribution.
Since the~DPE shape is fixed, the number of~DPE events~$N_{\text{DPE}}$ is the sole free parameter in the fit.
The DPE fit is first performed on bins with charge greater than the~DPE peak.
The resulting~DPE component is compared with the data, and bins in which it contributes~$>75\,\%$ of the content are selected.
A second fit restricted to these bins mitigates overestimation and improves the stability of the~DPE estimation. 
In figure~\ref{fig:PulseCharge}, the solid cyan line indicates the data range used for the fit, while the full~DPE component is depicted with a dash-dotted line.

The~SPE mean~$\mu_{\text{SPE}}$, standard deviation~$\sigma_{\text{SPE}}$, and the uncertainty of the~SPE mean charge are estimated directly from the data excluded from the~DPE fit.
By excluding data clearly associated with~DPE events, this approach provides a slightly improved estimate compared to simply taking the mean of the entire distribution.
The~SPE mean from pulse integrals, here~$0.85\,\text{pC}$, is systematically lower than that from trace integrals, here~$0.99\,\text{pC}$, because the pulse integral method does not fully capture the pulse area (Section~\ref{subsec:PulseCharacterization}).
The pulse charge spectrum is approximated via~KDE.
Lower and upper charge cuts for~SPE selection can be set using (i) the charge where the~KDE falls to~$50\,\%$ of its maximum and (ii) the~DPE peak, respectively. 
Here, the resulting~DPE contamination inside the chosen~SPE interval is~$\sim 2\,\%$, comparable to the~$\sim 1.4\,\%$ obtained from the trace charge model fit. 
Alternatively, data selection intervals can be defined in~PE units, e.g.,~$0.5$–$1.5$\,PE.

\subsection{Transit time spectrum}\label{subsec:TTSpec}
\begin{figure}[htb]
    \centering
    \includegraphics[width=1\linewidth]{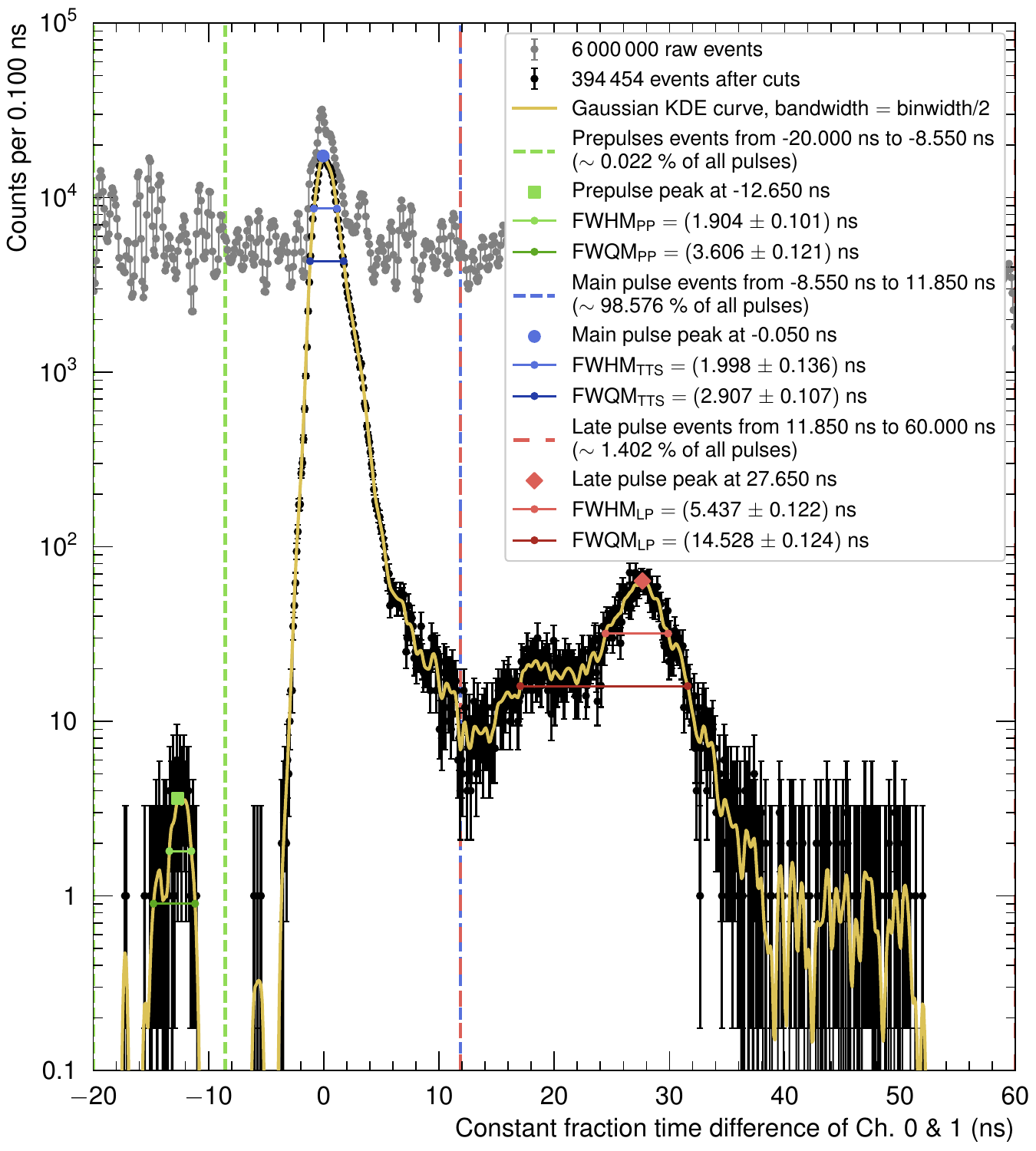}
    \caption{Example transit time distribution at~SPE intensity for a~9821B~PMT operated at~1800\,V. 
    The gold curve shows a Gaussian~KDE. 
    Vertical dashed lines delineate the prepulse, main pulse, and late pulse regions, and the markers indicate their peak positions. 
    Horizontal bars denote the~FWHM and~FWQM values summarized in the legend.
}
    \label{fig:TTS}
\end{figure}

The transit time spectrum is obtained from the time difference between the~PMT signal~(Ch.~0) and the~NIM logic signal~(Ch.~1), with an~SPE selection applied based on the pulse charge cut.
Because of cable/electronics delays and the laser~TTL-to-photon delay, the transit time spectrum has an arbitrary offset.
To ease discussions, we shift the spectrum so that the main peak is near zero, as figure~\ref{fig:TTS} shows. 

\begin{figure}[htb!]
    \centering
    \includegraphics[width=1\linewidth]{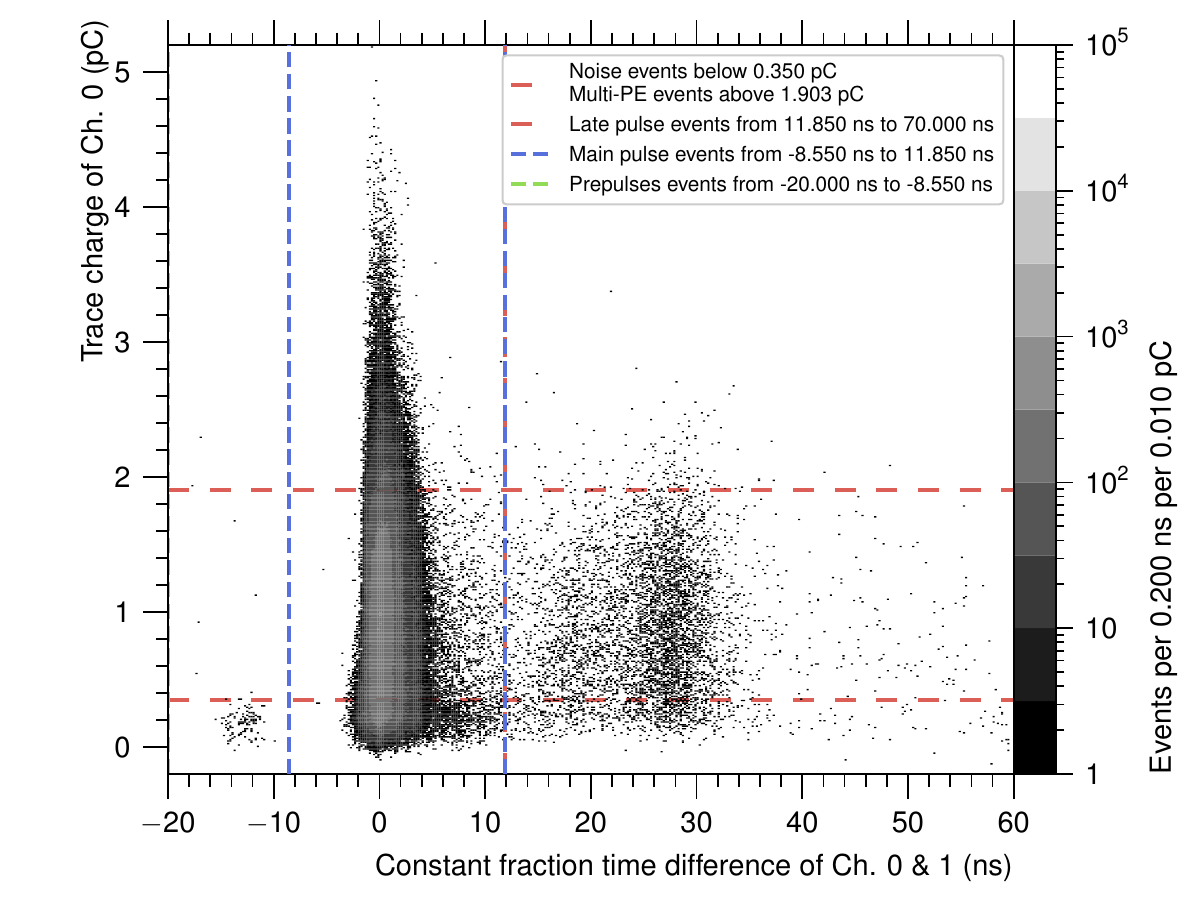}\\
    \includegraphics[width=1\linewidth]{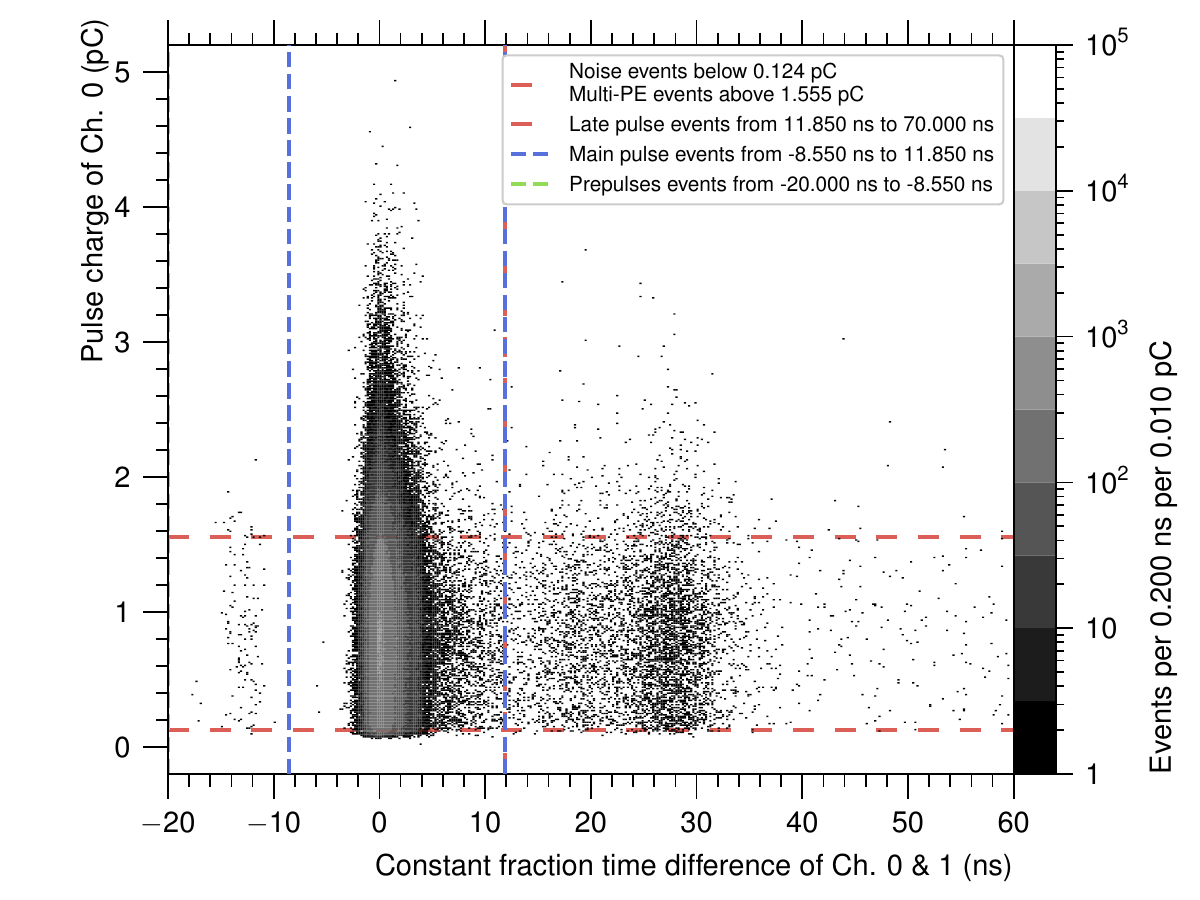}
    \caption{Charge-time correlations at low-light intensity using trace integral~(top) and pulse integral~(bottom) data.}
    \label{fig:2D_TI_TTS}
\end{figure}

A~KDE with appropriate bandwidth is used to smooth the distribution and to locate minima, which serve to separate pulse populations.
Following~\cite{PMT_SmiLomRan2004,TTandCh_SPE_R7081_PMTs_2012}, three populations are identified: prepulses~(PPs), main pulses, and late pulses~(LPs). 
Their fractions to the complete data set are counted in the~KDE-defined intervals, with an optional dark rate correction estimated from the constant pre-PP region if present.
Local full width at half maxima~(FWHMs) and full width at quarter maxima~(FWQMs) of the pulse populations are taken from the~KDE by linear interpolation, and uncertainties are propagated.
The three populations are described in the following:
\begin{itemize}
    \item \textbf{Prepulses} occur~$\Delta t_{\text{pre-main}} \simeq 12.6$\,ns before the main peak and contribute~$0.02\,\%$ of events.
    The~$\text{FWHM}_{\text{PP}} = (1.9 \pm 0.1)$\,ns is comparable to the main component's~$\text{FWHM}_{\text{TTS}} = (2.0 \pm 0.1)$\,ns.~PPs arise when photons pass the photocathode and strike the first dynode directly~\cite{PMT_prelate_2000}, yielding smaller charge because the first amplification stage is bypassed~\cite{TTandCh_SPE_R7081_PMTs_2012}. 
    Following~\cite{PMT_SmiLomRan2004}, adapting a photocathode-to-first-dynode distance of~$\sim2.5$\,cm (time-of-flight~$t_{\text{ToF}}\simeq 0.1$\,ns) and the measured pre-main delay~$\Delta t_{\text{pre-main}}$, the~PE drift time for the~9821B is~$t_{\text{PE\,drift}} = \Delta t_{\text{pre-main}} + t_{\text{ToF}} \simeq 12.7$\,ns.
    
    \item \textbf{Main pulses} are dominant~($98.58\,\%$) and occur from~$-8.5$\,ns to~$11.9$\,ns with~$\text{FWHM}_{\text{TTS}} = (2.0 \pm 0.1)$\,ns, which is close to the value in the~9821B datasheet~\cite{9821B_datasheet}. 
    The~TTS reflects variations in~PE paths/velocities and secondary-electron production, generally increasing with~PMT size. 
    Within the main component we distinguish~(i) early pulses from scattering that bypasses an amplification stage~(slightly lower charge),~(ii) regular pulses that mainly define the~TTS, and~(iii) delayed pulses from weaker acceleration near photocathode edges~\cite{TTandCh_SPE_R7081_PMTs_2012}.

    \item \textbf{Late pulses} comprise~$1.40\,\%$ of events, start at~$11.9$\,ns and peak at~$27.7$\,ns. 
    They originate from (in)elastic backscattering of photoelectrons at the first dynode.
    The longest delays are $\approx 2\,t_{\text{PE\,drift}} \simeq 25.4$\,ns, consistent with but slightly below the observed~LP peak-to-main pulse peak separation due to trajectory spread~\cite{PMT_SmiLomRan2004, TTandCh_SPE_R7081_PMTs_2012, PMT_Hama12inch_2013}. 
    Elastic backscatter tends to cluster near the~LP peak with charge similar to normal events, while inelastic cases arrive earlier with reduced charge~\cite{PMT_prelate_2000}. 
    A secondary bump broadens the~LP component, giving~$\text{FWHM}_{\text{LP}}=(5.4 \pm 0.1)$\,ns and $\text{FWQM}_{\text{LP}}=(14.5 \pm 0.1)$\,ns. 

\end{itemize}
Figure~\ref{fig:2D_TI_TTS} presents charge-time correlations using the trace integral~(top) and pulse integral~(bottom) methods. 
The trace integral method illustrates that prepulses are absent within the~SPE interval. 
Due to the systematically smaller charge of~PPs, integrating the full waveform, i.e., signal plus surrounding noise, masks their presence.

%--------------------------------------------------
\section{Results and discussion}\label{sec:results}
This section summarizes the performance of eight~9821B and two~9821QB PMTs from~ET Enterprises as well as one Hamamatsu~R9980~PMT assembly, characterized at~SPE intensity over a range of supply voltages. 
The “Q” designation denotes a version of the PMT equipped with a quartz glass window rather than a borosilicate glass window~\cite{ET_brochure}.
Although the R9980~PMT requires a negative bias, all voltages are quoted by magnitude.
All investigated~9821B devices are equipped with~B19A basic hardpin bases, MS75/90 mu-metal shields and~E638PFP-01 passive encapsulated voltage dividers.
Unless stated otherwise, measurements refer to room temperature and short signal cables in the light‑tight enclosure. 
Additional details are available in~\cite{PhD_MRStock2025}.

\subsection{Gain}\label{subsec:Gain_V}
\begin{figure}[htb!]
    \centering%0.6
    \includegraphics[width=0.99\linewidth]{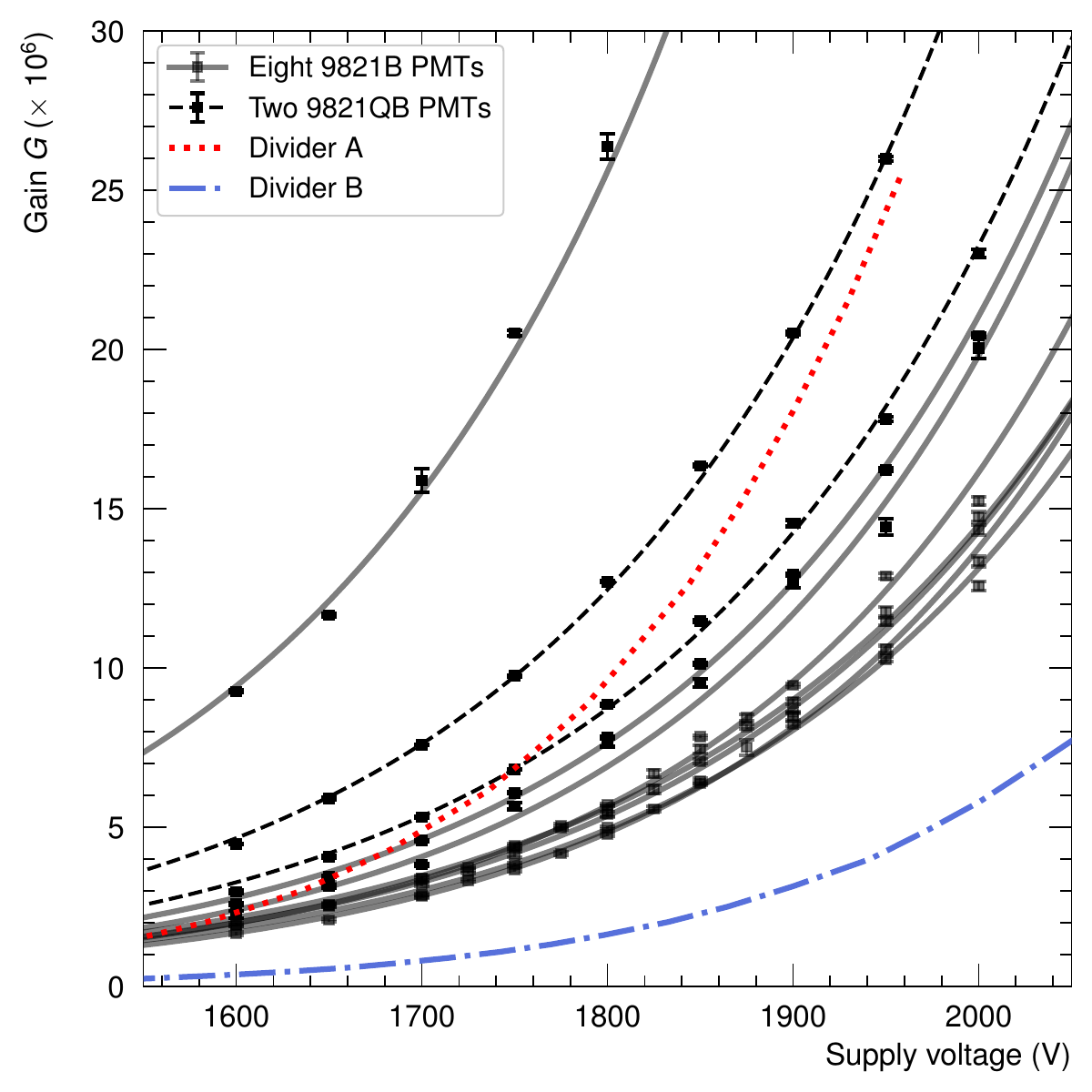}
    \includegraphics[width=0.99\linewidth]{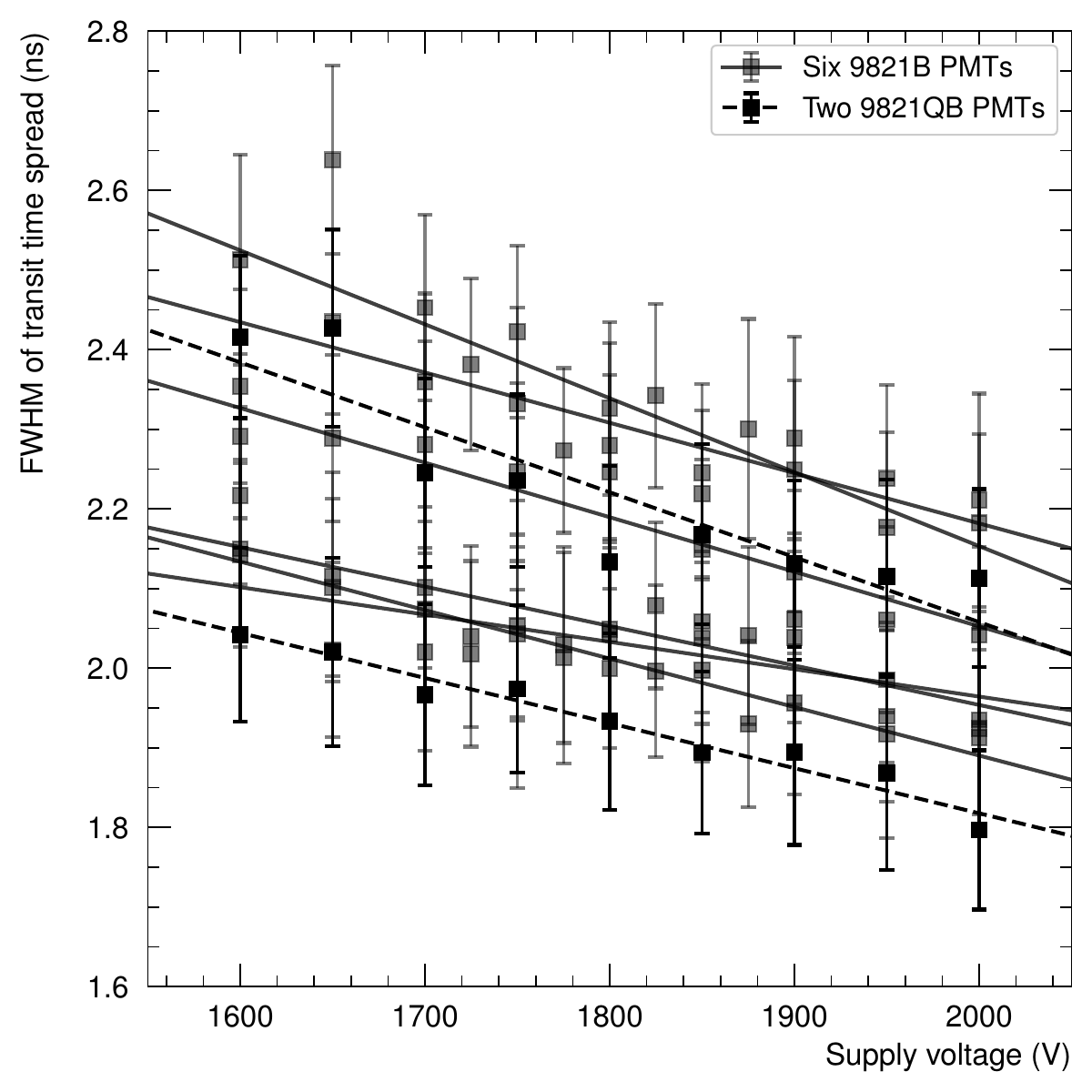}
    \caption{Gain~$G$ and timing~(TTS) versus supply voltage for~9821B (gray solid) and~9821QB~(black dashed)~PMTs. 
\\ \textbf{Top:}
The gain data was derived from the trace charge analysis and the curves are fits for each~PMT. 
The red dotted and blue dash-dotted curves indicate the expected gain scaling for divider~A and divider~B, respectively, whose data was taken from the~9821B datasheet~\cite{9821B_datasheet}. 
\textbf{Bottom:}
The~TTS decreases approximately linearly with voltage and lines are linear fits to each tube.}
    \label{fig:GainTTSVoltage}
\end{figure}

Figure~\ref{fig:GainTTSVoltage}~(top) shows the gain-voltage curves obtained from the trace charge analysis. 
For each device, an exponential model,~$G(U) = G_0 \exp{\left( U / U_0 \right)}$, where~$G_0$ and~$U_0$ are two fit parameters, describes the data well. 
Best‑fit curves follow the general scaling suggested by the manufacturer’s divider~A benchmark. 
Dividers~A and~B are alternative voltage divider configurations provided by the manufacturer for the same~PMT type. 
Divider~B employs a modified resistor distribution that supports higher pulsed anode currents and improved linearity compared to divider~A~\cite{9821B_datasheet}.
The parameters~$G_0$ and~$U_0$ vary tube‑to‑tube, emphasizing device‑specific behavior.
For orientation, the tubes typically exhibit~$U_0$ values of about~200\,V with gains larger than~$5 \times 10^6$ at~1800\,V.

Systematic uncertainties are dominated by the charge-integration methods, whereas the statistical uncertainty is almost negligible due to the large event statistics. 

\subsection{Peak‑to‑valley ratio~(P/V)}\label{subsec:P2V}
The peak-to-valley ratio~(P/V) increases with voltage for all tubes. 
For the~9821B/9821QB series,~P/V typically rises from about~1 to~2.3, while the~R9980 reaches~1~to~4.5 depending on the individual tube and noise conditions, which broaden the pedestal and reduce the~P/V.
For many voltages of some~PMTs, there was no~P/V.
The selected~ADC dynamic range setting required to sample large pulses at high voltages influences the~P/V via pedestal broadening as well.

\subsection{Timing}\label{subsec:TTS_V}
The~TTS decreases with voltage across the covered range for all tested tubes.
Figure~\ref{fig:GainTTSVoltage}~(bottom) shows the individual~FWHM of the~TTS together with linear fits illustrating this trend.
The measured values cluster around the~$\sim 2.2$\,ns quoted in the~9821B datasheet~\cite{9821B_datasheet}, but with clear device-specific variations.
The~TTS and its uncertainty arise from the sampling rate, the constant-fraction timing algorithm, the~KDE bandwidth and interpolation procedure for determining the~FWHM.

As detailed in~\cite{PhD_MRStock2025}, introducing an aperture in front of the photocathode reduces~TTS by suppressing the spread of photoelectron trajectories.
For detailed studies of the prepulse/late pulse components, it is recommended to record datasets containing several millions of~SPE events.

\subsection{Temperature and cabling}\label{subsec:Temperature}
Temperature scans from~$-50^\circ$C to~$+20^\circ$C were performed using long signal cables for several~9821(Q)B tubes and the~R9980.
Although the minimum recommended operating temperature for these models is~$-30^\circ$C~\cite{9821B_datasheet, R9980_datasheet}, all tubes remained operational down to~$-50^\circ$C. 

\begin{figure}[htb!]
    \centering
    \includegraphics[width=1\linewidth]{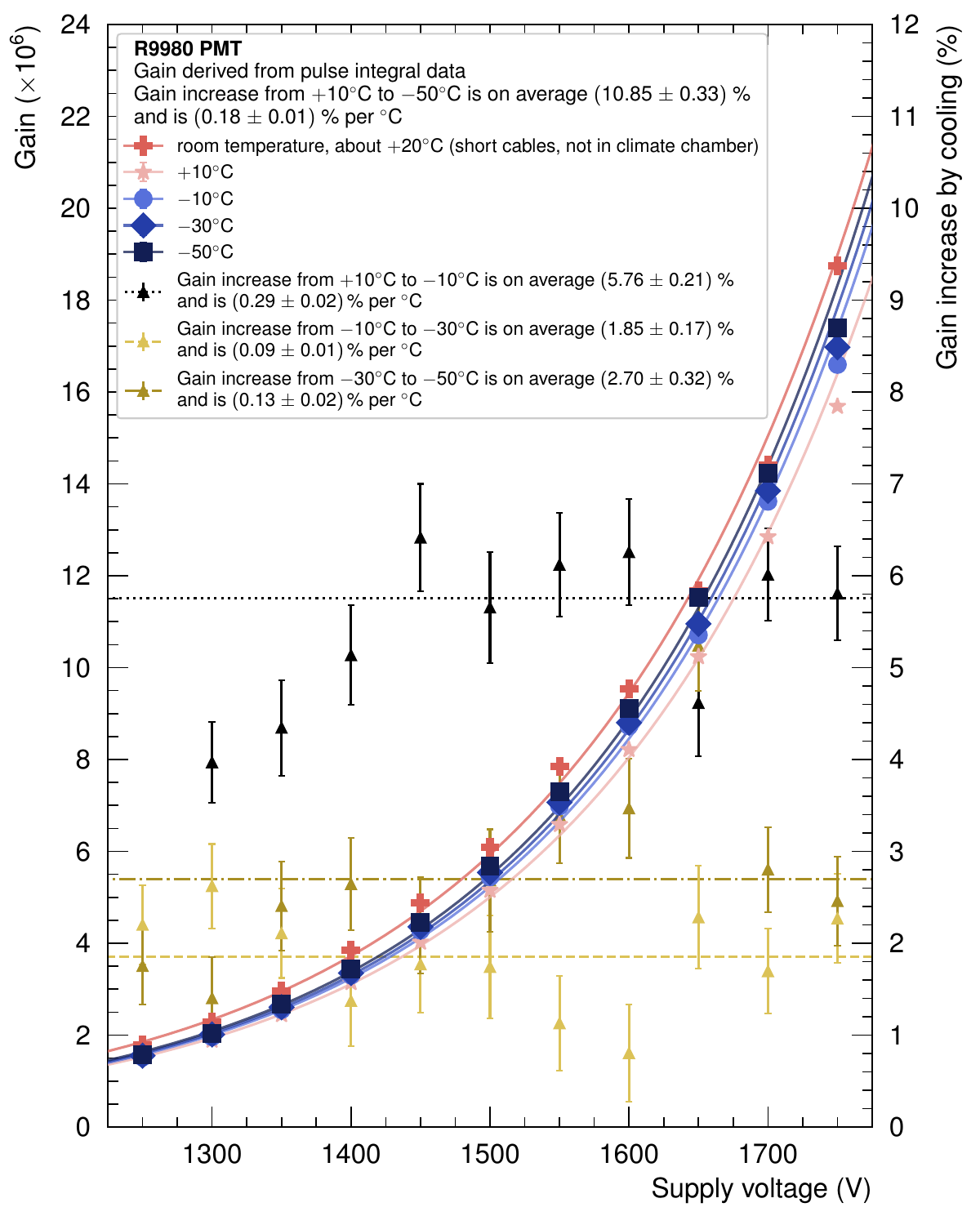}
    \caption{Gain-voltage curves of the~R9980~PMT measured at
    various temperatures. 
    The gain increases with decreasing temperature and with shorter signal cables, the latter explaining the comparatively high room-temperature values measured in the light-tight enclosure.}
    \label{fig:R_PI_Temp}
\end{figure}

\begin{figure}[htb!]
    \centering
    \includegraphics[width=1\linewidth]{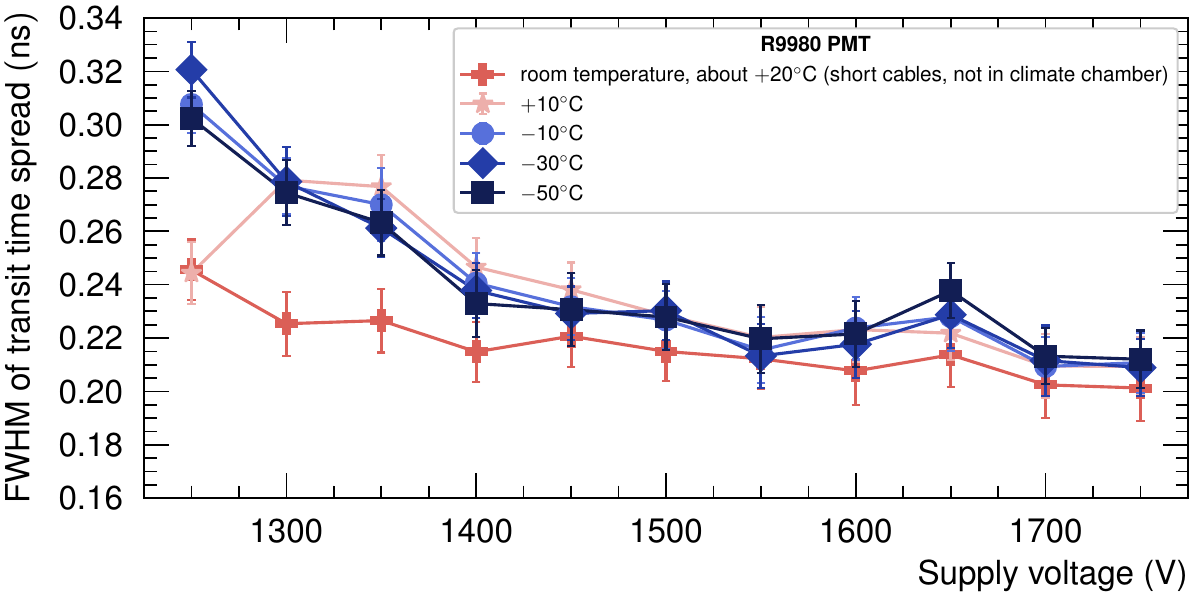}
    \caption{FWHM of the~TTS for the~R9980 decreases with increasing supply voltage.
    The smaller~TTS values at~$+20^\circ$C result from the use of short cables, while the temperature-controlled data used long cables.
    No significant temperature dependence is observed.}
    \label{fig:R_TTS_Temp}
\end{figure}

The gain increases systematically toward lower temperature by a few tenths of a percent per degree Celsius.
This behavior is also tube-dependent and can be explained by changes of the surface properties of the dynodes, affecting the secondary emission factor and thus leading to higher gains~\cite{LeoBook}.
Additionally, decreased resistance of the~PMT divider circuit and signal cables results in higher charge output~\cite{QualiPMTs_XENON_2017}.
Figure~\ref{fig:R_PI_Temp} shows the~R9980 gain derived from pulse integrals as a function of temperature and voltage.
Within uncertainties no robust temperature dependence is observed for the~TTS, as illustrated in figure~\ref{fig:R_TTS_Temp}. 
All measured~TTS values are compatible with the~0.2\,ns specification for the~R9980~\cite{R9980_datasheet}. 
The~P/V ratio shows no clear temperature trend.
A systematic dependence on cable length is observed.
Shorter cables yield slightly higher gain and a smaller apparent~TTS due to reduced attenuation and dispersion. 
This effect exceeds the thermal influence and must be accounted for when comparing different setups.

%--------------------------------------------------
\section{Conclusion}\label{sec:conclusion}
A compact, table-top setup for photosensor characterization has been presented and validated with ET~Enterprises~9821B/9821QB tubes and a Hamamatsu~R9980 assembly.
Picosecond laser excitation combined with waveform-based analysis enables reproducible measurements of gain, peak-to-valley ratio~(P/V), transit time spread~(TTS), and prepulse/late pulse populations under well-controlled laboratory conditions, including optional temperature scans.

The analysis introduces a model-independent, data-driven method to extract the double photoelectron~(DPE) fraction directly from the pulse charge spectrum, allowing robust contamination estimates without reliance on parametric charge models.
All tested tubes follow the expected exponential gain-voltage dependence and exhibit clear device-to-device variations.
The~TTS decreases with increasing voltage and reaches values compatible with the specifications, again with noticeable device-to-device variation.
Late pulses occur at the percent level, while prepulses remain at the sub-percent level with reduced charge.
Cooling increases the gain by a few tenths of a percent per degree Celsius, whereas no significant temperature dependence is observed for the~TTS or~P/V in the investigated range.
Cable length has a measurable influence on both gain and apparent~TTS, underscoring the need for consistent electrical routing when comparing different setups or operating conditions.

The characterization procedures developed here provide the basis for describing the~PMT responses in our liquid scintillator~(LS)~R\&D program and are used to verify and validate the performance of the associated~LS test setups.
The presented methods therefore supply a reproducible and practice-oriented framework for~SPE-level charge and timing studies that can be readily adapted to other~PMT types.
Beyond the specific devices studied, the setup and methodology are, in principle, transferable to other photosensors with~SPE sensitivity and can be extended to additional observables~(e.g.\ afterpulsing, high-light linearity, or photon detection efficiency).
Such extensions could, for example, make use of an additional reference photosensor, such as a~SiPM operated in parallel to the~PMT and viewing the same laser pulses, to constrain the absolute transit time and to verify the high-intensity response and linearity.
In this way, the setup offers a reliable reference for laboratory-scale photosensor characterization.

\section*{CRediT authorship contribution statement}
\textbf{M.\,R.\,S.}: Conceptualization, Data curation, Formal analysis, Investigation, Methodology, Software, Validation, Visualization, Writing -- original draft.
\textbf{H.\,Th.\,J.\,S.}: Conceptualization, Investigation, Methodology, Resources, Validation.
\textbf{U.\,F.}: Investigation, Software, Validation, Writing -- review \& editing.
\textbf{L.\,S.}: Data curation, Investigation, Validation.
\textbf{L.\,O.}: Conceptualization, Funding acquisition, Methodology, Project administration, Resources, Supervision, Writing -- review \& editing.

\section*{Declaration of interest}
The authors declare that they have no known competing financial interests or personal relationships that could have appeared to influence the work reported in this paper.

%---------------- Acknowledgments and AI-use declaration ----------------
\section*{Acknowledgments}
We thank Korbinian Stangler and Florian Kübelbäck for assistance with laser-beam imaging during setup verification. 
We also acknowledge the Detector Lab of the Excellence Cluster PRISMA$^+$ at Johannes Gutenberg University Mainz for providing access to the climate chamber and making it available at TUM for measurements.
We thank Tobias Sterr at the University of Tübingen for performing initial tests of several photomultiplier tubes using their laser test facility.

\section*{Data availability}
The datasets underlying the current study are available from the corresponding author
on reasonable request.

\section*{Code availability}
The analysis code used in this work is available from the corresponding author on
reasonable request.

\section*{Acronyms}

\begin{tabular}{ll}
\textbf{ADC} & Analog-to-digital converter \\
\textbf{DAQ} & Data acquisition \\
\textbf{DPE} & Double photoelectron \\
\textbf{EMG} & Exponentially modified Gaussian distribution \\
\textbf{EMGH} & Heaviside modified \\
 & exponentially modified Gaussian distribution \\
\textbf{FWHM} & Full width at half maximum \\
\textbf{FWQM} & Full width at quarter maximum \\
\textbf{GH} & Heaviside modified Gaussian distribution  \\
\textbf{HV} & High voltage \\
\textbf{JUNO} & Jiangmen Underground Neutrino Observatory \\
\textbf{KDE} & Kernel density estimation \\
\textbf{LENA} & Low Energy Neutrino Astronomy \\
\textbf{LP} & Late pulse \\
\textbf{LS} & Liquid scintillator \\
\textbf{NIM} & Nuclear Instrumentation Modules \\
\textbf{PDF} & Probability density function \\
\textbf{PE} & Photoelectron \\
\textbf{PP} & Prepulse \\
\textbf{P/V} & Peak-to-valley ratio \\
\textbf{PMT} & Photomultiplier tube \\
\textbf{QE} & Quantum efficiency \\
\textbf{R\&D} & Research and development \\
\textbf{SiPM} & Silicon photomultiplier \\
\textbf{SPE} & Single photoelectron \\
\textbf{TPE} & Triple photoelectron \\
\textbf{TTL} & Transistor-transistor logic \\
\textbf{TTS} & Transit time spread \\
\end{tabular}

\FloatBarrier
%---------------- Appendix ----------------
%\appendix

%---------------- References ----------------
\bibliographystyle{elsarticle-num}
\bibliography{references}

@phdthesis{PhD_MRStock2025,
  author       = {{M.\,R. Stock}},
  title        = {{Development and Characterization of Novel Liquid Scintillators for the Detection of Neutrinos in Future Large-Scale Detectors}},
  school       = {Technical University of Munich},
  year         = {2025},
  type         = {{Ph.D. thesis}},
  note = {\url{https://mediatum.ub.tum.de/1782355}}
}

@article{QualiPMTs_DC_2011,
  author       = {{C. Bauer et al.}},
  title        = {{Qualification tests of 474 photomultiplier tubes for the inner detector of the {Double Chooz} experiment}},
  journal      = {JINST},
  volume       = {\textbf{6}},
  pages        = {P06008},
  year         = {2011},
  doi          = {10.1088/1748-0221/6/06/P06008}
}

@article{QualiPMTs_XENON_2017,
  author       = {{P. Barrow et al.}},
  title        = {{Qualification tests of the {R11410-21} photomultiplier tubes for the {XENON1T} detector}},
  journal      = {JINST},
  volume       = {\textbf{12}},
  pages        = {P01024},
  year         = {2017},
  doi          = {10.1088/1748-0221/12/01/P01024}
}

@article{TTandCh_SPE_R7081_PMTs_2012,
  author       = {{F. Kaether, C. Langbrandtner}},
  title        = {{Transit time and charge correlations of single photoelectron events in {R7081} photomultiplier tubes}},
  journal      = {JINST},
  volume       = {\textbf{7}},
  pages        = {P09002},
  year         = {2012},
  doi          = {10.1088/1748-0221/7/09/P09002}
}

@article{BRIGATTI2005521,
  author       = {{A. Brigatti, A. Ianni, P. Lombardi, G. Ranucci, O.\,Ju. Smirnov}},
  title        = {{The photomultiplier tube testing facility for the {Borexino} experiment at {LNGS}}},
  journal      = {Nucl. Instrum. Methods Phys. Res. A},
  volume       = {\textbf{537}},
  number = {3},
  pages        = {521--536},
  year         = {2005},
  doi          = {10.1016/j.nima.2004.07.248}
}

@article{PMT_SmiLomRan2004,
  author       = {{O.\,Ju. Smirnov, P. Lombardi, G. Ranucci}},
  title        = {{Precision Measurements of Time Characteristics of {ETL9351} Photomultipliers}},
  journal      = {Instrum. Exp. Tech.},
  volume       = {\textbf{47}},
  pages        = {69--80},
  year         = {2004},
  doi          = {10.1023/B:INET.0000017255.60520.e0}
}

@article{SCHWARZ201964,
  author  = {{M. Schwarz et al.}},
  title   = {Measurements of the lifetime of orthopositronium in the {LAB}-based liquid scintillator of {JUNO}},
  journal = {Nucl. Instrum. Methods Phys. Res. A},
  volume  = {\textbf{922}},
  pages   = {64--70},
  year    = {2019},
  doi     = {10.1016/j.nima.2018.12.068}
}

@article{Diwan2020,
	author = {{M.\,V. Diwan}},
	title = {Statistics of the charge spectrum of photo-multipliers and methods for absolute calibration},
	journal = {JINST},
    volume = {\textbf{15}},
    pages = {P02001},
	year = {2020},
	doi =  {10.1088/1748-0221/15/02/p02001}
}

@article{RDossi2000,
	author = {{R. Dossi, A. Ianni, G. Ranucci, O.\,Ju. Smirnov}},
	title = {Methods for precise photoelectron counting with photomultipliers},
	journal = {Nucl. Instr. and Meth. in Phys. Res. Sec. A},
    volume = {\textbf{451}},
    number = {3},
    pages = {623--637},
	year = {2000},
	doi = {10.1016/S0168-9002(00)00337-5}
}

@article{PMT_DEAP3600_2019,
   author={{P.-A. Amaudruz et al.}},
   title = {{In-situ characterization of the Hamamatsu R5912-HQE photomultiplier tubes used in the DEAP-3600 experiment}},
   journal = {Nucl. Instr. and Meth. in Phys. Res. Sec. A},
   volume = {\textbf{922}},
   doi = {10.1016/j.nima.2018.12.058},
   pages = {373-384},
   year = {2019},
}

@article{PMT_prelate_2000,
	title = {{Studies of prepulses and late pulses in the 8" electron tubes series of photomultipliers}},
	author = {{B.\,K. Lubsandorzhiev, P.\,G. Pokhil, R.\,V. Vasiljev, A.\,G. Wright}},
    journal = {Nucl. Instr. and Meth. in Phys. Res. Sec. A},
    volume = {\textbf{442}},
    number = {1},
    pages = {452--458},
    year = {2000},
    issn = {0168-9002},
    doi = {10.1016/S0168-9002(99)01272-3}
}

@article{PMT_Hama12inch_2013,
    author = {{J. Brack et al.}},
    title = {{Characterization of the Hamamatsu R11780 12 in. photomultiplier tube}},
	journal = {{Nucl. Instr. and Meth. in Phys. Res. Sec. A}},
    volume = {\textbf{712}},
    pages = {162--173},
    year = {2013},
    issn = {0168-9002},
    doi = {10.1016/j.nima.2013.02.022}
}

@article{PMT_Bellamy_1994,
    author = {{E.\,H. Bellamy et al.}},
    title = {Absolute calibration and monitoring of a spectrometric channel using a photomultiplier},
	journal = {{Nucl. Instr. and Meth. in Phys. Res. Sec. A}},
    volume = {\textbf{339}},
    number = {3},
    pages = {468-476},
    year = {1994},
    doi = {10.1016/0168-9002(94)90183-X}
}

@book{LeoBook,
	author        = {{W.\,R. Leo}},
	title         = {{Techniques for Nuclear and Particle Physics Experiments: A How-to Approach; 2nd ed.}},
	publisher     = "Springer",
	address       = "Berlin",
	year          = "1994",
	doi           = {10.1007/978-3-642-57920-2}
}

@misc{R9980_datasheet,
	author = {{Hamamatsu Photonics}},
	title = {{{Photomultiplier tube R9980}}},
	note = {\url{https://www.hamamatsu.com/content/dam/hamamatsu-photonics/sites/documents/99_SALES_LIBRARY/etd/R9980_TPMH1341E.pdf}}
}

@misc{9821B_datasheet,
	author = {{ET Enterprises}},
	title = {{{78 mm (3") photomultiplier, 9821B series data sheet}}},
	note = {\url{https://et-enterprises.com/images/data_sheets/9821B.pdf}}
}

@misc{iminuit,
  author={{H. Dembinski et al.}},
  title={scikit-hep/iminuit (v2.29.1)},
  publisher    = {Zenodo},
  year         = {2024},
  doi          = {10.5281/zenodo.13371653}
}

@article{scipy,
  author  = {{P. Virtanen et al. {(SciPy 1.0 Contributors)}}},
  title   = {{{SciPy} 1.0: Fundamental Algorithms for Scientific
            Computing in Python}},
  journal = {Nature Methods},
  year    = {2020},
  volume  = {\textbf{17}},
  pages   = {261--272},
  adsurl  = {https://rdcu.be/b08Wh},
  doi     = {10.1038/s41592-019-0686-2}
}

@software{root,
  author       = {{F. Rademakers et al.}},
  title        = {root-project/root: v6.20/04},
  year         = {2020},
  publisher    = {Zenodo},
  version      = {(v6-20-04)},
  doi          = {10.5281/zenodo.3895855},
}

@article{Sympy,
     title = {{SymPy: symbolic computing in Python}},
     author = {{A. Meurer et al.}},
     year = {2017},
     volume = {\textbf{3}},
     pages = {e103},
     journal = {PeerJ Computer Science},
     doi = {10.7717/peerj-cs.103}
}

@misc{uproot,
  author       = {{J. Pivarski et al.}},
  title        = {Uproot (v5.3.13)},
  year         = {2024},
  publisher    = {Zenodo},
  doi          = {10.5281/zenodo.13821023}
}

@article{numpy,
 title         = {{Array programming with {NumPy}}},
 author        = {{C.\,R. Harris et al.}},
 year          = {2020},
 journal       = {Nature},
 volume        = {\textbf{585}},
 number        = {7825},
 pages         = {357--362},
 doi           = {10.1038/s41586-020-2649-2},
 publisher     = {Springer Science and Business Media {LLC}}
}

@article{matplotlib,
  author    = {{J.\,D. Hunter}},
  title     = {{Matplotlib: A 2D graphics environment}},
  journal   = {{Computing in Science \& Engineering}},
  volume    = {\textbf{9}},
  number    = {3},
  pages     = {90--95},
  publisher = {IEEE COMPUTER SOC},
  doi       = {10.1109/MCSE.2007.55},
  year      = {2007}
}

@misc{pandas,
  author       = {{The pandas development team}},
  title        = {pandas-dev/pandas: Pandas (v2.0.3)},
  year         = {2023},
  publisher    = {Zenodo},
  doi          = {10.5281/zenodo.8092754}
}

@book{python,
     author = {{G. Van Rossum, F.\,L. Drake}},
     title = {{Python 3 Reference Manual}},
     year = {2009},
     isbn = {1441412697},
     publisher = {CreateSpace},
     address = {Scotts Valley, CA}
}

@phdthesis{PhDTippmann,
	author = {{M.\,A. Tippmann}},
	title = {{Comprehensive photosensor research for large liquid scintillator neutrino detectors}},
	year = {2021},
	school = {Technische Universität München},
	pages = {855},
	note = {\url{https://mediatum.ub.tum.de/1611529}}
}

@article{WbLS,
    doi = {10.1088/1748-0221/19/09/P09008},
    year = {2024},
    volume = {\textbf{19}},
    number = {09},
    pages = {P09008},
    author = {{H.\,Th.\,J. Steiger et al.}},
    title = {{Development, characterization and production of a novel water-based liquid scintillator based on the Surfactant TRITON™ X-100}},
    journal = {JINST}
}

@article{SlowLS,
    doi = {10.1088/1748-0221/19/09/P09015},
    year = {2024},
    volume = {\textbf{19}},
    number = {09},
    pages = {P09015},
    author = {{H.\,Th.\,J. Steiger et al.}},
    title = {{Development of a bi-solvent liquid scintillator with slow light emission}},
    journal = {JINST}
}

@article{JUNO,
    doi = {10.1140/epjc/s10052-022-11002-8},
    year = {2022},
    volume = {\textbf{82}},
    number = {1168},
    author = {{A. Abusleme et al. (JUNO Collaboration)}},
    title = {{Mass testing and characterization of 20-inch PMTs for JUNO}},
    journal = {Eur. Phys. J. C}
}

@article{TAUP2023,
    doi = {10.22323/1.441.0287},
    year = {2024},
    author = {{M.\,R. Stock et al.}},
    title = {{Scintillation Time Profiles of Slow Organic and Water-Based Liquid Scintillators using a Pulsed Neutron Beam}},
    journal = {PoS(TAUP2023)287}
}

@misc{ET_brochure,
    author = {{ET Enterprises}},
    title = {{Photomultipliers}},
    note = {\url{https://et-enterprises.com/images/brochures/Pmt_Brochure.pdf}}
}

\end{document}